\documentclass[12pt]{article}
\usepackage{amsfonts}
\usepackage{amssymb}

\def\hybrid{\topmargin -20pt    \oddsidemargin 0pt
        \headheight 0pt \headsep 0pt
        \textwidth 6.35in       
        \textheight 9.25in       
        \marginparwidth .875in
        \parskip 5pt plus 1pt   \jot = 1.5ex}

\hybrid

\def\baselinestretch{1.2}

\catcode`\@=11

\def\marginnote#1{}
\newcount\hour
\newcount\minute
\newtoks\amorpm
\hour=\time\divide\hour by60
\minute=\time{\multiply\hour by60 \global\advance\minute by-\hour}
\edef\standardtime{{\ifnum\hour<12 \global\amorpm={am}%
        \else\global\amorpm={pm}\advance\hour by-12 \fi
        \ifnum\hour=0 \hour=12 \fi
        \number\hour:\ifnum\minute<10 0\fi\number\minute\the\amorpm}}
\edef\militarytime{\number\hour:\ifnum\minute<10 0\fi\number\minute}

\def\draftlabel#1{{\@bsphack\if@filesw {\let\thepage\relax
   \xdef\@gtempa{\write\@auxout{\string
      \newlabel{#1}{{\@currentlabel}{\thepage}}}}}\@gtempa
   \if@nobreak \ifvmode\nobreak\fi\fi\fi\@esphack}
        \gdef\@eqnlabel{#1}}
\def\@eqnlabel{}
\def\@vacuum{}
\def\draftmarginnote#1{\marginpar{\raggedright\scriptsize\tt#1}}

\def\draft{\oddsidemargin -.5truein
        \def\@oddfoot{\sl preliminary draft \hfil
        \rm\thepage\hfil\sl\today\quad\militarytime}
        \let\@evenfoot\@oddfoot \overfullrule 3pt
        \let\label=\draftlabel
        \let\marginnote=\draftmarginnote
   \def\@eqnnum{(\theequation)\rlap{\kern\marginparsep\tt\@eqnlabel}%
\global\let\@eqnlabel\@vacuum}  }

\def\preprint{\twocolumn\sloppy\flushbottom\parindent 2em
        \leftmargini 2em\leftmarginv .5em\leftmarginvi .5em
        \oddsidemargin -.5in    \evensidemargin -.5in
        \columnsep .4in \footheight 0pt
        \textwidth 10.in        \topmargin  -.4in
        \headheight 12pt \topskip .4in
        \textheight 6.9in \footskip 0pt
        \def\@oddhead{\thepage\hfil\addtocounter{page}{1}\thepage}
        \let\@evenhead\@oddhead \def\@oddfoot{} \def\@evenfoot{} }

\def\numberbysection{\@addtoreset{equation}{section}
        \def\theequation{\thesection.\arabic{equation}}}

\def\underline#1{\relax\ifmmode\@@underline#1\else
        $\@@underline{\hbox{#1}}$\relax\fi}

\def\titlepage{\@restonecolfalse\if@twocolumn\@restonecoltrue\onecolumn
     \else \newpage \fi \thispagestyle{empty}\c@page\z@
        \def\thefootnote{\fnsymbol{footnote}} }

\def\endtitlepage{\if@restonecol\twocolumn \else \newpage \fi
        \def\thefootnote{\arabic{footnote}}
        \setcounter{footnote}{0}}  

\catcode`@=12
\relax

\def\figcap{\section*{Figure Captions\markboth
        {FIGURECAPTIONS}{FIGURECAPTIONS}}\list
        {Figure \arabic{enumi}:\hfill}{\settowidth\labelwidth{Figure
999:}
        \leftmargin\labelwidth
        \advance\leftmargin\labelsep\usecounter{enumi}}}
 \relax
\def\tablecap{\section*{Table Captions\markboth
        {TABLECAPTIONS}{TABLECAPTIONS}}\list
        {Table \arabic{enumi}:\hfill}{\settowidth\labelwidth{Table
999:}
        \leftmargin\labelwidth
        \advance\leftmargin\labelsep\usecounter{enumi}}}
 \relax
\def\reflist{\section*{References\markboth
        {REFLIST}{REFLIST}}\list
        {[\arabic{enumi}]\hfill}{\settowidth\labelwidth{[999]}
        \leftmargin\labelwidth
        \advance\leftmargin\labelsep\usecounter{enumi}}}
 \relax
\makeatletter
\newcounter{pubctr}
\def\publist{\@ifnextchar[{\@publist}{\@@publist}}
\def\@publist[#1]{\list
        {[\arabic{pubctr}]\hfill}{\settowidth\labelwidth{[999]}
        \leftmargin\labelwidth
        \advance\leftmargin\labelsep
        \@nmbrlisttrue\def\@listctr{pubctr}
        \setcounter{pubctr}{#1}\addtocounter{pubctr}{-1}}}
\def\@@publist{\list
        {[\arabic{pubctr}]\hfill}{\settowidth\labelwidth{[999]}
        \leftmargin\labelwidth
        \advance\leftmargin\labelsep
        \@nmbrlisttrue\def\@listctr{pubctr}}}
 \relax
\makeatother
\newskip\humongous \humongous=0pt plus 1000pt minus 1000pt

\newif\ifdtup

\relax

\def\be{\begin{equation}}
\def\ee{\end{equation}}
\def\ba{\begin{eqnarray}}
\def\ea{\end{eqnarray}}

\def\del{\partial}

\def\k{\kappa}
\def\r{\rho}
\def\a{\alpha}

\def\b{\beta}

\def\G{\Gamma}

\def\D{\Delta}
\def\e{\epsilon}

\def\th{\theta}

\def\m{\mu}
\def\n{\nu}
\def\om{\omega}
\def\Om{\Omega}

\def\s{\sigma}

\def\cN{{\cal N}}

\def\no{\noindent}

\def\qq{\qquad}

\def\IR{\relax{\rm I\kern-.18em R}}
\renewcommand{\theequation}{\thesection.\arabic{equation}}
\csname @addtoreset\endcsname{equation}{section}

\def \ha {{1\over 2}}

\def \ov {\over}

\def\IR{\relax{\rm I\kern-.18em R}}
\def\inv{^{\raise.15ex\hbox{${\scriptscriptstyle -}$}\kern-.05em 1}}

\begin{document}


\newcommand{\beq}{\begin{equation}}
\newcommand{\eeq}[1]{\label{#1}\end{equation}}
\newcommand{\ber}{\begin{eqnarray}}
\newcommand{\eer}[1]{\label{#1}\end{eqnarray}}
\newcommand{\eqn}[1]{(\ref{#1})}
\begin{titlepage}
\begin{center}

\vskip -.1 cm
\hfill hep--th/0507169\\

\vskip 1in

{\Large \bf Supersymmetric solutions based on $Y^{p,q}$ and $L^{p,q,r}$}

\vskip 0.4in

{\bf Konstadinos Sfetsos} \phantom{x}and\phantom{x} {\bf Dimitrios
Zoakos} \vskip 0.1in

\vskip .2in

Department of Engineering Sciences, University of Patras\\
26110 Patras, Greece\\
{\footnotesize{\tt sfetsos@des.upatras.gr, dzoakos@upatras.gr}}\\

\end{center}

\vskip .3in

\centerline{\bf Abstract}

\no We explicitly realize supersymmetric cones based on the
five-dimensional $Y^{p,q}$ and $L^{p,q,r}$ Einstein--Sasaki
spaces. We use them to construct supersymmetric type-IIB
supergravity solutions representing a stack of D3- and D5-branes
as warped products of the six-dimensional cones and $\IR^{1,3}$.

\noindent

\vskip .4in
\noindent

\end{titlepage}
\vfill
\eject

\def\baselinestretch{1.2}


\baselineskip 20pt


\section{Introduction}

In recent works the number of explicit examples of known
five dimensional Einstein--Sasaki
metrics was considerably enlarged
by a new class of such metrics interpolating in a certain sense
between the round $S^5$ sphere and the $T^{1,1}$
space \cite{Gauntlett:0403002}.
These are of cohomogeneity one, with principal
orbits $SU(2)\times U(1)\times U(1)$ and in order to
satisfy global regularity issues their parametric space
is characterized by two coprime positive integers $p$
and $q$. Hence, these were called $Y^{p,q}$ spaces.
This class has been further generalized by taking the
BPS limits of Euclideanized Kerr--de Sitter black hole metrics with
two independent angular momenta parameters
\cite{Cvetic:0504225}. This construction
leads to local Einstein--Sasaki metrics of cohomogeneity two, with
$U(1)\times U(1)\times U(1)$ principal orbits. Similarly, these metrics,
called $L^{p,q,r}$, are characterized by positive coprime integers
$p$, $q$ and $r$ in order that they smoothly extend onto complete,
non-singular compact
manifolds. The $Y^{p,q}$ spaces come as special limits of
the $L^{p,q,r}$ ones when the angular parameters coincide and a $U(1)$
symmetry factor gets enhanced into $SU(2)$. Similarly, the $T^{1,1}$ space
results by a further symmetry enhancement.

One advantage of having explicit five-dimensional regular spaces is that
they can be used as a base in the construction of six-dimensional Ricci-flat
cones, which in turn are basic blocks for the ten-dimensional
supergravity solutions representing the gravitational field of stacks of
branes and the dual description
of supersymmetric gauge theories within the gauge/gravity
correspondence \cite{Maldacena:9711200}-\cite{Witten:9802150}.
The usual cone one constructs suffers from a singularity in its tip
and therefore part of the effort is to regularize it. A basic example is the
six-dimensional cone based on the $T^{1,1}$ space in which the conical
singularity were first smoothened out in the so called deformed and
resolved conifolds \cite{Candelas}, by introducing a parameter,
and keeping finite at the tip of the cone either an $S^2$ or an $S^3$ factor.
In addition, there is also the regularized conifold in which the original
curvature singularity becomes a removable bolt singularity \cite{PZ}.
Introducing D3-branes and taken into account their backreaction
transforms the
Ricci-flat solution of the cone times the Minkowski space
into a warped solution of the
full type-IIB supergravity \cite{Keh}-\cite{Klebanov:2000hb}.
We note that having a regular six-dimensional cone
does not necessarily imply the
regularity of the ten-dimensional solution
(see, in particular, \cite{PZ} that emphasizes that).
The purpose of this paper is to construct the six-dimensional supersymmetric
cones based on the newly discovered $Y^{p,q}$ and $L^{p,q,r}$ spaces
and use them for the construction of the ten-dimensional type-IIB
supergravity solutions that include the brane backreaction.

This letter is organized as follows: In section 2 we present a
brief review of the relevant aspects of the $Y^{p,q}$ and
$L^{p,q,r}$ spaces. In section 3 we explicitly construct
supersymmetric six-dimensional cone solutions based on these
spaces. They depend on a constant moduli parameter as in the
regularized conifold. In section 4 we construct supersymmetric
supergravity solutions of a stack of D3- and D5-branes on these
cones within type-IIB supergravity. They have the expected
behaviour in the UV, but still suffer from a singularity in the
IR.


\section{Brief review of the $Y^{p,q}$ and $L^{p,q,r}$ spaces}

In this section we provide a short review of some relevant to our construction
aspects of the $Y^{p,q}$ and $L^{p,q,r}$ spaces and also
comment on their relation. For details the reader should really consult the
literature.

\subsection{$Y^{p,q}$ geometry}

The five dimensional $Y^{p,q}$ geometry in its canonical form is
described by the following metric
\cite{Gauntlett:0402153}-\cite{Martelli:0411238}
\be
ds^2_5=ds^2_4+\left({1 \ov 3}d\psi+\s\right)^2\ ,
\label{sf}
\ee
where the four dimensional metric is
\be
ds^2_4 ={1-cy \ov 6}\left(d\th^2+\sin^2\th d\phi^2\right)+{dy^2
\ov w(y)q(y)}+{1 \ov 36}w(y)q(y)(d\b+c\cos\th d\phi)^2\ ,
\label{Ypq}
\ee
with
\be
\s= -{1\ov 3} \cos\th d\phi + {1\ov 3} y (d\b+c \cos\th d\phi)\ .
\ee
and
\ba
w(y)={2(a-y^2) \ov
1-cy} \ , \quad q(y)={a-3y^2+2cy^3 \ov a-y^2}\ .
\ea
Therefore, it can be seen as a $U(1)$ bundle over a four-dimensional
Einstein-K\"{a}hler metric with the K\"{a}hler two-form
given by $d\s=2J_4$. It can be checked explicitly that the
four-dimensional metric is Einstein with $R_{\m\n}=6g_{\m\n}$ and hence
the five-dimensional metric is Einstein--Sasaki with
$R_{\m\n}=4g_{\m\n}$.
The coordinate $y$ ranges between the two smallest roots of the
cubic equation $a-3y^2+2cy^3=0$, so that the signature of the metric remains
Euclidean. These are given in terms of the
coprime integers $p$ and $q$ with explicit expressions that don't concerns us
here. In order also to obtain a compact manifold the coordinate $\a$ has a
finite range. The remaining ones $\th$, $\phi$ and
$\psi$ have periods $\pi$, $2\pi$ and $2\pi$, respectively.


\subsection{$L^{p,q,r}$ geometry}

The five-dimensional $L^{p,q,r}$ geometry is described by the
following metric
\cite{Cvetic:0504225,Martelli:0505027}
\ba
ds^2_5=ds_4^2+ \left(d\tau+\s\right)^2\ ,
\label{5d}
\ea
where the four-dimensional metric is
\ba
ds^2_4&=&{\r^2dx^2 \ov 4\D_x}+{\r^2d\th^2 \ov \D_{\th}}+{\D_x \ov
\r^2}\left({\sin^2\th \ov \a}d\phi+{\cos^2\th \ov
\b}d\psi\right)^2
\nonumber\\
&&+{\D_{\th}\sin^2\th\cos^2\th \ov
\r^2}\left[(1-x/\a)d\phi-(1-x/\b)d\psi\right]^2
\ea
and
\ba
&& \s=(1-x/\a)\sin^2\th \ d\phi+(1-x/\b)\cos^2\th \ d\psi \ , \qq
\r^2=\D_{\th}-x \ ,
\nonumber\\
&& \D_x=x(\a-x)(\b-x)-\m \ , \qq
\D_{\th}=\a\cos^2\th+\b\sin^2\th \ .
\ea
The five-dimensional metric has the standard form,
as in the $Y^{p,q}$ case. The
parameter $\mu$ is trivial and can be set to any non-zero
constant by rescaling $\a$, $\b$, and $x$, hence the metric depend on
two parameters.
The principal orbits $U(1)\times U(1)\times U(1)$ of the metric
degenerate at $\theta=0$ and $\theta= {\pi \ov 2}$ and at the
roots of the cubic function $\Delta_x$. In order to obtain metrics
on non-singular manifolds the ranges of the coordinates should be
$0<\th<{\pi \ov 2}$ and $x_1<x<x_2$, where $x_1$ and $x_2$ are the
two smallest roots of the equation $\D_x=0$. The ranges of the
coordinates $\phi$ and $\psi$ are determined using the notion of
``surface gravity'', important in black hole solutions of Lorentzian signature.
The analysis of the behavior at each collapsing orbit can be
realized by examining the associated Killing vector $\ell$ whose
length vanishes at the degeneration surface. By normalizing the
Killing vector so that its ``surface gravity'' $\kappa$ is equal
to unity, one obtains a translation generator $\partial/\partial
\chi$, where $\chi$ is a local coordinate near the degeneration
surface. The metric extends smoothly onto the surface if
$\chi$ has period $2\pi$. The ``surface gravity'' is
\be
\k^2={g^{\m\n}(\partial_{\m}\ell^2)(\partial_{\n}\ell^2) \ov
4\ell^2} \ ,
\label{surf.gr}
\ee in the limit the degeneration surface is reached.
At the degeneration surfaces $\th=0$ and $\th={\pi \ov
2}$ the normalized killing vectors are $\partial/\partial \phi$
and $\partial/\partial \psi$ respectively, so the periodicity of
the coordinates $\phi$ and $\psi$ is 0 to $2\pi$. At the
degeneration surfaces $x=x_1$ and $x=x_2$, the
associated normalized Killing vectors $\ell_1$ and $\ell_2$ are
given by
\be
\ell_i = c_i\,{\del \ov \del \tau} + a_i\, {\del \ov
\del\phi}+b_i\,{\del \ov \del\psi}\ ,
\label{ells}
\ee where the constants $c_i$, $a_i$ and $b_i$ are given by
\ba
a_i={\a c_i \ov
x_i - \a}\ ,\qq b_i={\b c_i \ov x_i-\b}\ ,\qq c_i =
{(\a-x_i)(\b-x_i) \ov 2(\a+\b)x_i -\a\b-3x_i^2}\ .
\ea
Similarly to the case of $Y^{p,q}$, all parameters are eventually given in
terms of three
coprime positive integers $p$, $q$ and $r$ so that the manifolds are
complete and free of singularities.


\subsubsection{Connection with $Y^{p,q}$}

If one sets $p+q=2r$, implying $\a=\b$,
the metric \eqn{5d} reduce to \eqn{Ypq} with $Y^{p,q}=L^{p-q,p+q,p}$.
Then the relation of variables and parameters is given by
\cite{Butti:0505220}
\ba
x\rightarrow{\a \ov 3}(1+2cy)\ , \quad
\th\rightarrow{\th \ov 2}, \quad \phi-\psi\rightarrow-\phi\ , \quad
\phi+\psi\rightarrow{\b \ov c}\ , \quad
3\tau+\phi+\psi\rightarrow-\psi\
\label{coord.trans}
\ea
and
\be \mu={4 \ov 27}(1-ac^2)\a^3.
\ee
After the coordinate transformation (\ref{coord.trans}) the
Killing vectors for the degeneration surfaces $\th=0$ and
$\th=\pi$ are $(-\partial/\partial \phi-\partial/\partial
\psi+c\partial/\partial\b)$ and $(\partial/\partial
\phi-\partial/\partial \psi+c\partial/\partial\b)$, respectively.
At the degeneration surfaces $y=y_1$ and $y=y_2$, where $y_1$ and
$y_2$ are the roots of the equation $q(y)=0$, the normalized
killing vectors $\ell_1$ and $\ell_2$ are given by
\be
\ell_i={\partial \ov \partial \psi}-{1 \ov y_i}{\partial \ov
\partial \b}\ , \qq i=1,2
\label{ell'}
\ee
and correspond to the vectors in \eqn{ells}.


\section{The six-dimensional cones}

In this section we explicitly solve the supersymmetric Killing
spinor equations and determine the six-dimensional cones. The
latter are by construction Ricci-flat.

\subsection{The cone over the $Y^{p,q}$ space}

First we construct a six-dimensional supersymmetric cone over the
$Y_{p,q}$ space as a base. The metric ansatz is
\ba
ds^2_6&=&dr^2+A(r)^2\left({1 \ov
3}d\psi+\s\right)^2+B(r)^2ds_4^2\ .
\label{6dYcone}
\ea
We will use the vielbein basis
\ba
&& e^1=B(r)\sqrt{{1-cy \ov 6}}d\th\ , \qq
e^2=B(r)\sqrt{{1-cy \ov 6}}\sin\th d\phi\ ,
\\
&& e^3=B(r){dy \ov \sqrt{w(y)q(y)}}\ , \qq e^4=B(r){1 \ov
6}\sqrt{w(y)q(y)}(d\b+c\cos\th d\phi)  \ ,
\nonumber\\
&& e^5=A(r){1 \ov
3}\left[d\psi-\cos\th d\phi+y(d\b+c\cos\th d\phi)\right]\ , \qq
e^6=dr\ .
\nonumber
\ea
The non-vanishing components of the spin connection are
\ba
&& \om^{12}=-{1 \ov
B}\left[\cot\th \left({6 \ov 1-cy}\right)^{1/2}e^2+{A \ov B}e^5-
{c \ov 2(1-cy)}\sqrt{w q}\ e^4 \right]  \ ,
\nonumber\\
&&\om^{34}=-{1 \ov B}\left[{\partial \ov \partial y}\sqrt{w
q}\ e^4+{A \ov B}e^5\right] \ ,
\nonumber\\
&&\om^{14}={1 \ov B}{c \ov 2(1-cy)}\sqrt{w q}\ e^2\ , \qq
\om^{15}=-{A \ov B^2}e^2\ ,
\nonumber\\
&& \om^{13}=-{1 \ov B}{c \ov 2(1-cy)}\sqrt{w q}\ e^1\ , \qq
\om^{25}={A \ov B}e^1\ ,
\\
&& \om^{24}=-{1 \ov B}{c \ov 2(1-cy)}\sqrt{w q}\ e^1\ ,
\qq
 \om^{45}={A \ov B^2}e^3\ ,
\nonumber\\
&& \om^{23}=-{1 \ov B}{c \ov 2(1-cy)}\sqrt{w q}\ e^2\ ,\qq
\om^{35}=-{A \ov B^2}e^4\ ,
\nonumber\\
&& \om^{i6}={B' \ov B}e^i, \quad i=1\ldots4 \ , \qq \om^{56}={A'
\ov A}e^5 \ , \nonumber \ea where prime denotes differentiation
with respect to $r$.\footnote{One might try a more general ansatz
than \eqn{6dYcone} by putting different functions of $r$ in front
of every vielbein. However, it turns that the consistent with
supersymmetry solution in the end simplifies the ansatz to that in
\eqn{6dYcone}. This is consistent with the observation of
\cite{Franco:0502113} that, generically the $Y^{p,q}$ manifolds do
not admit complex deformations.} The Killing spinor equation are
\be
\del_\mu \e + {1\ov 4} \om^{ab}_\mu \G_{ab}\e=0\ .
\ee
In
analyzing this set of equations we found necessary to impose the
two projections
\be
\G_{12}\e=\G_{34}\e=-\G_{56}\e\ ,
\label{prki}
\ee
hence
reducing supersymmetry to $1/4$ of the maximal. The Killing spinor
turns out to be
\be
\e=e^{{1 \ov 2}\psi \G_{12}}\e_0\ .
\label{heh}
\ee
In addition we obtained the following
system of differential equations that determine the functions
$A(r)$ and $B(r)$
\ba
B' = {A \ov B} \ ,\qq  A' = 3-2{A^2 \ov
B^2}\ .
\label{jdh3}
\ea
The general solution to the system is
\be
B^2=R^2\ ,\qq A^2=R^2 \left(1+{C \ov R^6}\right)\ ,
\ee
where $C$
is a constant. The relation of the two variables r and R is via
the differential \be dr=\left(1+{C \ov R^6}\right)^{-1/2} dR \ .
\ee Note that we have absorbed a second integration constant by a
suitable redefinition of the variable $R$. After substituting the
solution of the killing spinor equations to (\ref{6dYcone}) the
metric takes the simple form\footnote{This solution belongs to the
class of examples considered in \cite{bergery,Page} by
solving the second order field equations. We thank C. Pope for the
information. A form of this solution was also obtained in
\cite{Pal:0501012} but without any claim or proof on
supersymmetry.}
\be
ds_6^2=\left(1+{C \ov
R^6}\right)^{-1}dR^2+R^2\left(1+{C \ov R^6}\right)\left({1 \ov
3}d\psi+\s \right)^2+R^2ds^2_4 \ . \label{6dY.sol}
\ee
We have
checked that this metric has the same killing vectors with
(\ref{Ypq}), with degeneration surfaces $\th=0$, $\th=\pi$,
$y=y_1$ and $y=y_2$.

\no
The asymptotic behavior for large values of $R$ takes the
universal form
\be
ds_6^2\simeq dR^2+R^2ds_5^2, \quad {\rm as}
\quad R\rightarrow\infty
\label{infYR}
\ee
and it describes the
usual cone whose base is given by the five dimensional metric
(\ref{Ypq}). This solution is exact for all values of $R$
since it can be obtained by simply
letting $C=0$.
The constant $C$ changes the solution drastically towards the interior.
When $C\geq 0$, the variable $R\geq 0$ and then the
manifold has a curvature singularity at $R=0$. However, if $C=-a^6<0$, where $a$
is a real positive constant, then the variable $R\geq a$.
To examine the behaviour of the metric near $R=a$ we
change into the new radial variable $t=\sqrt{6a(R-a)}$. We find
\ba
ds_6^2\simeq a^2ds^2_4+{1 \ov 9}dt^2+t^2\left({1
\ov 3}d\psi+\s\right)^2 \quad {\rm as} \quad t\rightarrow 0\ .
\label{bolt}
\ea
Therefore, near $t=0$ and for constant $y$,
$\th$, $\b$ and $\phi$, the metric behaves (up to $1/9$) as $dt^2+t^2d\psi^2$
which shows that $t=0$ is a {\it bolt} singularity \cite{Gibbons}
which is removable since the periodicity of the angle
is $0\leq \psi <2\pi$. The full solution
interpolates between \eqn{bolt} for $R\rightarrow a$ and
(\ref{infYR}) for $R\rightarrow\infty$.
This is similar to that found in \cite{hersfe} for the
cones over the symmetric coset spaces $SU(2)^n/U(1)^{n-1}$ that includes
the regularization of the singular conifold on $T^{1,1}$ for $n=2$
\cite{PZ}. However, in our case we do not have a completely non-singular
solution at the supergvavity level.\footnote{We thank C. Pope for
a correspondence on this.} The Einstein--Kahler four-dimensional base is singular.
At best it has orbifold singularities, when $4p^2-3q^2=n^2$,
where $n\in Z$.  The $Y^{p,q}$
metrics are then an orbifold $U(1)$ bundle over this Einstein--Kahler base
orbifold \cite{Martelli:0411238}. Nevertheless,
string theory has probably more success with orbifold singularities
than true curvature singularities since in some cases the singularity
is resolved before the "smoothening" \cite{dixon}.
It is interesting to investigate this further.


\subsection{The cone over the $L^{p,q,r}$ space}

To construct the six-dimensional supersymmetric cone
over $L_{p,q,r}$ we make the ansatz
\ba
ds^2_6&=&dr^2+A(r)^2\left(d\tau+\s\right)^2+B(r)^2ds_4^2\
\label{6dcone}
\ea
and use the vielbein basis
\ba
&& e^1=B(r){\r \ov
\D_{\th}^{1/2}}d\th\ , \qq e^2=B(r){\D_{\th}^{1/2}\sin\th\cos\th
\ov \r} \left({\a-x \ov \a}d\phi-{\b-x \ov \b}d\psi\right)\ ,
\nonumber\\
&& e^3=B(r){\D_x^{1/2} \ov \r}\left({\sin^2\th
\ov \a}d\phi+{\cos^2\th \ov \b}d\psi\right)\ , \qq e^4=B(r){\r
\ov 2\D_x^{1/2}}dx \,
\label{vielbein}\\
&& e^5=A(r)\left(d\tau+\s\right)\ ,\qq  e^6=dr\ .
\nonumber\
\ea
After some tedious algebra we found that the non-vanishing components
of the spin connection are
\ba
&& \om^{12}=-{1 \ov B}\left[
\left(2\cot2\th{\D_{\th}^{1/2} \ov \r}-{\a-\b \ov
2}\sin2\th\left({1 \ov \r\D_{\th}^{1/2}}-{\D_{\th}^{1/2} \ov
\r^3}\right)\right)e^2 +{\D_{x}^{1/2} \ov
\r^3}e^3+{A \ov B}e^5\right] \ ,
 \nonumber\\
&& \om^{34}={1 \ov B}\left[\left({3x^2-2(\a+\b)x+\a\b\ \ov
\r\D_{x}^{1/2}}+{\D_x^{1/2} \ov \r^3} \right)e^3-{A \ov
B}e^5+{\a-\b \ov 2}\sin2\th{\D_{\th}^{1/2} \ov
\r^3}e^2\right] \ ,
\nonumber\\
&& \om^{14}=-{1 \ov B}\left[{\D_x^{1/2} \ov \r^3}e^1-{\a-\b \ov
2}\sin2\th{\D_{\th}^{1/2} \ov \r^3}e^4\right]\ , \qq
\om^{15}=-{A \ov B^2}e^2\ ,
\nonumber\\
&& \om^{13}=-{1 \ov B}\left[{\D_x^{1/2} \ov \r^3}e^2-{\a-\b \ov
2}\sin2\th{\D_{\th}^{1/2} \ov \r^3}e^3\right]\ , \qq \om^{25}={A
\ov B^2}e^1\ ,
\\
&& \om^{24}=-{1 \ov B}\left[{\D_x^{1/2} \ov
\r^3}e^2-{\a-\b \ov 2}\sin2\th{\D_{\th}^{1/2} \ov
\r^3}e^3\right],\qq
 \om^{45}={A \ov B^2}e^3\ ,
\nonumber\\
&& \om^{23}=+{1 \ov B}\left[{\D_x^{1/2} \ov \r^3}e^1-{\a-\b \ov
2}\sin2\th{\D_{\th}^{1/2} \ov \r^3}e^4\right]\ ,
\qq \om^{35}=-{A \ov B^2}e^4\ , \nonumber\\
&& \om^{i6}={B' \ov B}e^i\ , \qq i=1\ldots4\ , \qq
\om^{56}={A' \ov A}e^5 \ .
\nonumber
\ea
The set of projections obtained by analyzing the Killing spinor
equations are the same as in the case of the
cone over the $Y^{p,q}$ space in \eqn{prki} and similarly the system of
differential equations \eqn{jdh3} determining the functions $A(r)$ and $B(r)$.
The Killing spinor is
\be
\e=e^{{1 \ov 2}(3\tau+\phi+\psi)\G_{12}}\e_0\ .
\ee
Finally the solution takes the simple form
\be
ds_6^2=\left(1+{C \ov
R^6}\right)^{-1}dR^2+R^2ds_4^2+R^2\left(1+{C \ov
R^6}\right)(d\tau+\s)^2\ .
\label{6d.sol}
\ee
The metric above has the same killing vectors with (\ref{5d}),
with degeneration
surfaces $\th=0$, $\th=\pi/2$, $x=x_1$ and $x=x_2$.
As in the case of the cone over $Y^{p,q}$, for
$C=-a^6$ the metric \eqn{6d.sol} is free of curvature singularities, but it
has the singularities associated with the four-dimensional Einstein--Kahler space.


\section{\bf Warped type-IIB solutions}

In order to construct ten-dimensional supersymmetric warped solutions
we utilize the procedure developed
in \cite{Grana:0009211,Gubser:0010010}.
We will use the cone over the $L^{p,q,r}$ space.\footnote{For
related work with the usual cone over $L^{p,q,r}$ see also
\cite{Martelli:0505027} and \cite{Gepner:0505039}.}
This procedure was also recently used to construct a solution
with the usual cone over $Y^{p,q}$ \cite{Herzog:0412193}.
The first step is to find a harmonic $(2,1)$ form $\Om_{2,1}$. For this
reason we will use the local K\"{a}hler form $J_4$ on the
K\"{a}hler-Einstein base
\ba
 J_4 & = & \tilde{e}^1\wedge\tilde{e}^2+\tilde{e}^3\wedge\tilde{e}^4
\nonumber\\
& = &
\sin\th \cos\th d\th\wedge\left[(1-x/\a)d\phi-(1-x/\b) d\psi\right]
\\
&& -{1 \ov 2}dx\wedge\left(1/\a \sin^2\th
\ d\phi+1/\b \cos^2\th\ d\psi\right)\ ,
\nonumber
\ea
where $\tilde e^i=e^i/A(r)$.
In turn, based on
\cite{Franco:0402120}, it is possible to
construct a $\Om_{2,1}$ form from a (1,1) form $\om$ such that $*_4\om=-\om$,
$d\om=0$ and $\om\wedge J_4=0$.
Such a form is similar to the
one proposed in \cite{Herzog:0412193} and \cite{Martelli:0505027}
for the case of the usual cone on the $Y^{p,q}$ and $L^{p,q,r}$,
respectively. We have explicitly that
\ba
\om & = &
{1\ov \rho^4} (\tilde{e}^1\wedge\tilde{e}^2-\tilde{e}^3\wedge\tilde{e}^4)
\nonumber\\
& = &  {1\ov \rho^4}\Big[
\sin\th \cos\th\ d\th \wedge \left((1-x/\a)d\phi -(1-x/\b)d\psi\right)
\label{om1}\\
&&
 +\ha dx\wedge
(1/\a \sin^2\th\ d\phi +1/\b\cos^2\th\ d\psi)\Big]\ ,
\nonumber
\ea
where the overall factor in the first line has been fixed by demanding that
$d\om=0$.\footnote{
A second possibility is
\ba
\om  =  {1 \ov \sin2\th(\D_{\th}\D_x)^{1/2}}
(\tilde{e}^1\wedge\tilde{e}^4-\tilde{e}^2\wedge\tilde{e}^3)
 =  -{1\ov 2 \a\b}\ d\phi \wedge d\psi + {\rho^2\ov 2 \sin 2\th
\Delta_\th \Delta_x}\ d\th\wedge dx\ .
\label{om2}
\ea
However, this form is
singular at $\th=0, \pi$ and $x=x_1,x_2$ and cannot
be used to construct a complex 3-form with
well defined associated charges. We thank C. Herzog for
a correspondence on this.}
In order to check that the above form is indeed
$(1,1)$ we introduce the set of complex coordinates
(This should be equivalent to that presented in \cite{Martelli:0505027} in a
different coordinate system)
\ba
&& \eta_1=-{\cot\th \ov \D_{\th}}d\th+{\b-x \ov 2\D_x}dx+{i \ov
\a}d\phi\ ,
\nonumber\\
&& \eta_2={\tan\th \ov \D_{\th}}d\th+{\a-x
\ov 2\D_x}dx+{i \ov \b}d\psi\ ,
\label{jsd9}\\
&& \eta_3=\left(1-{a^6 \ov
R^6}\right)^{-1}{dR \ov R}+i\tilde{e}^5
-\eta_1(\a-x)\sin^2\th-\eta_2(\b-x)\cos^2\th\ .
\nonumber
\ea
It can be shown that the $\eta_i$'s indeed are closed and by construction
$(1,0)$ forms. Using \eqn{jsd9} we can solve for $d\th, dx,d\phi$ and $d\psi$
in terms of $\eta_{1,2}$ and their complex conjugates. Then after
substituting into \eqn{om1} (and \eqn{om2} for that matter) and some
algebra we may show that both expressions indeed represent (1,1) forms.

Next we construct a (2,1) form as the wedge product of a (1,0)
form and $\om$
\be
\Om_{2,1}=K\left[\left(1-{a^6 \ov
R^6}\right)^{-1}{dR \ov R}+i\tilde{e}^5\right]\wedge \om\ ,
\ee
where K is a normalization constant. It is easily verified that the
$\Om_{2,1}$ form is closed and imaginary self-dual in the six
dimensional space, namely
\be
d\Om_{2,1}=0\  , \qq *_6\Om_{2,1}=i\Om_{2,1} \ .
\ee
For the supergravity solution, we take the real RR
$F_3$ and NSNS $H_3$ forms to be
\be
iM\Om_{2,1}=F_3+{i \ov
g_s}H_3
\ee
and therefore
\be
F_3=-MK\tilde{e}^5\wedge \om\ , \qq
H_3=g_sMK\left(1+{C \ov R^6}\right)^{-1}{dR \ov R}\wedge \om\ ,
\ee
where M is another normalization constant.
The ansatz for the warped metric of the ten-dimensional type-IIB solution is
\be
ds_{10}^2=H^{-1/2}ds^2_4+H^{1/2}\left[\left(1-{a^6 \ov
R^6}\right)^{-1}dR^2+R^2ds_4^2+R^2\left(1-{a^6 \ov
R^6}\right)(d\tau+\s)^2\right]\ ,
\label{10d}
\ee
where the warp factor $H$ in generally depends on $R$, $x$ and $\th$.
There is no dilaton or axion field,
while the self-dual five form is
\be
g_sF_5=d(H^{-1})\wedge d^4x+*_{10}\left[d(H^{-1})\wedge d^4x\right] \ ,
\ee
which after some algebra takes the form
\ba
g_s F_5&=&-{H^{-2}}\left({\partial H \ov
\partial R}dR+{\partial H \ov
\partial x}dx+{\partial H \ov
\partial \th}d\th\right)\wedge d^4x
\nonumber\\
&&-{\partial H \ov
\partial R}R^5\left(1-{a^6 \ov
R^6}\right){\sin2\th \ov 4\a\b}\r^2d\tau\wedge d\th\wedge dx\wedge
d\phi\wedge d\psi \nonumber\\
&&-{\partial H \ov
\partial x}R^3{\sin2\th \ov
\a\b}\D_x d\tau\wedge dR\wedge d\th\wedge d\phi\wedge d\psi \\
&&+{\partial H \ov
\partial \th}R^3{\sin2\th \ov
4\a\b}\D_{\th} d\tau\wedge dR\wedge dx\wedge d\phi\wedge
d\psi\ .
\nonumber
\ea
To determine the warped factor we substitute in the Bianchi identity
\be
dF_5=H_3\wedge F_3
\ee
and obtain a second order partial differential equation
whose precise form depends on which one of \eqn{om1}
or \eqn{om2} we use to construct the 3-forms $H_3$ and $F_3$.
If we use (\ref{om1}) in the Bianchi identity we obtain
\ba
&&{1\ov R^3} {\partial \ov
\partial R}\left({\partial H \ov
\partial R}R^5\left(1-a^6/R^6\right)\right) + {4\ov \rho^2} {\partial \ov
\partial x}\left({\partial H \ov \partial x}\D_x\right)
\nonumber\\
&&+ {1/\rho^2\ov \sin2\th} {\partial \ov
\partial \th}\left({\partial H \ov
\partial \th}\sin2\th\D_{\th}\right) =
-2 {g_s^2M^2K^2 \ov \r^8 R^4}\left(1-a^6/ R^6\right)^{-1} \ .
\label{PDE1}
\ea
In the special case with $\a=\b$ we can check that this equations indeed
reduces to that in \cite{Herzog:0412193}
after we also make the consistent assumption that $H$ is $\th$-independent.
We were not able to find exact solutions of \eqn{PDE1} in the generic case.
In that respect note that it is not consistent to assume $\th$-independence of
the solutions. Perhaps the work of \cite{Kihara:2005nt} who study the
Laplacian in the $Y^{p,q}$ spaces will be useful in
that direction as well.
Nevertheless, we may easily see that for large $R$ it
exhibits the generic behaviour as $H\sim \ln R/R^4$.
Towards the infrared
for $R\to a$ it is seen that there is a singularity
since $H\sim \ln^2 (R-a)$.

Perhaps the most important open issue concerns the construction of a
supergravity solution utilizing the $Y^{p,q}$ and $L^{p,q,r}$ spaces and
being dual to $\cN=1$ gauge theories, in which the IR singularity is smoothened
out. Let's recall that some times a useful approach in constructing
supersymmetric spaces representing cones with smoothened out singularities
is via gauged supergravities. In particular, many such solutions having
an $SU(2)$ isometry
were found using the eight-dimensional supergravity of \cite{Salam}
resulting from
dimensionally reducing the eleven-dimensional supergravity of \cite{Cremmer}
(see in particular the works \cite{EN}-\cite{Barcelona1}\cite{hersfe}).
The use of the lower dimensional gauged supergravity disentangles certain
technical issues which are
due to the complexity of the base manifolds (in our case see the
expressions for the spin connections in section 3). We believe that at least
for the case of
solutions having the $Y^{p,q}$ manifold as an internal part
the use of gauged supergravity could be proven quite useful.


\newpage

\centerline {\bf Acknowledgments}

\no
We would like to thank C.N.Pope and D. Martelli
for a correspondence and A. Paredes for a
discussion.
We acknowledge the financial support provided through the European
Community's program ``Fundamental Forces and Symmetries of the
Universe" with contract MRTN-CT-2004-005104, the INTAS contract
03-51-6346 ``Strings, branes and higher-spin gauge fields'', as
well as by the Greek Ministry of Education program
$\Pi$Y$\Theta$A$\Gamma$OPA$\Sigma$ with contract 89194. In
addition D.Z. acknowledges the financial support provided through
the Research Committee of the University of Patras for a
``K.Karatheodory'' fellowship under contract number 3022.
We also thank Ecole Polytechnique and CERN for hospitality and financial
support during part of this work.


\renewcommand{\theequation}{\thesection.\arabic{equation}}
\csname @addtoreset\endcsname{equation}{section}


\end{document}